\documentclass[10pt,twocolumn,letterpaper]{article}

\usepackage{iccv}
\usepackage{times}
\usepackage{epsfig}
\usepackage{graphicx}
\usepackage{amsmath}
\usepackage{amssymb}
\usepackage[font=small]{caption}

\usepackage{color}
\usepackage{algorithm}
\usepackage{algpseudocode}
\usepackage{multirow}
\usepackage[accsupp]{axessibility}  
\DeclareMathOperator*{\argmax}{argmax}

\usepackage[pagebackref=true,breaklinks=true,letterpaper=true,colorlinks,bookmarks=false]{hyperref}

\iccvfinalcopy 

\def\iccvPaperID{4117} 

\ificcvfinal\fi

\begin{document}

\title{ReconfigISP: Reconfigurable Camera Image Processing Pipeline}


\author{
	Ke Yu$^{1,2}$ \hspace{9pt} Zexian Li$^{1,3}$ \hspace{9pt} Yue Peng$^{1}$ \hspace{9pt} Chen Change Loy$^{4}$ \hspace{9pt} Jinwei Gu$^{1,5}$\\
	$^{1}$\small{SenseTime Research and Tetras.AI} \hspace{13pt}
	$^{2}$\small{CUHK-SenseTime Joint Lab, The Chinese University of Hong Kong} \\ 
	$^{3}$\small{Beihang University} \hspace{13pt}
	$^{4}$\small{S-Lab, Nanyang Technological University} \hspace{13pt}
	$^{5}$\small{Shanghai AI Laboratory}\\
	{\tt\small ericyu16@hotmail.com \hspace{5pt} lizexian0427@gmail.com} \\
	{\tt\small \{pengyue, gujinwei\}@tetras.ai \hspace{5pt} ccloy@ntu.edu.sg}
}

\maketitle
\ificcvfinal\thispagestyle{empty}\fi
\thispagestyle{empty}
\begin{abstract}
	\vspace{-0.2cm}
	Image Signal Processor (ISP) is a crucial component in digital cameras that transforms sensor signals into images for us to perceive and understand. Existing ISP designs always adopt a fixed architecture, \eg, several sequential modules connected in a rigid order. Such a fixed ISP architecture may be suboptimal for real-world applications, where camera sensors, scenes and tasks are diverse. In this study, we propose a novel Reconfigurable ISP (ReconfigISP) whose architecture and parameters can be automatically tailored to specific data and tasks. In particular, we implement several ISP modules, and enable backpropagation for each module by training a differentiable proxy, hence allowing us to leverage the popular differentiable neural architecture search and effectively search for the optimal ISP architecture. A proxy tuning mechanism is adopted to maintain the accuracy of proxy networks in all cases. Extensive experiments conducted on image restoration and object detection, with different sensors, light conditions and efficiency constraints, validate the effectiveness of ReconfigISP. Only hundreds of parameters need tuning for every task.\footnote{Codes will be available at \url{https://www.mmlab-ntu.com/project/reconfigisp/}}
	\vspace{-0.3cm}
\end{abstract}

\section{Introduction}

A digital camera uses an ISP to transform original RAW images to high-quality RGB images that can be displayed on screen. Existing ISP systems typically adopt a human-designed pipeline~\cite{karaimer2016software} comprised of specialized modules, each of which addresses a subtask, such as denoising~\cite{guo2019toward,plotz2018neural}, demosaicing~\cite{gharbi2016deep,khashabi2014joint}, white balancing~\cite{hsu2008light}, \etc. 
Traditional ISP is highly modular and efficient. However, once the pipeline is fixed, it can hardly be adapted to different application scenarios. A costly and manual tuning of parameters is required.

\begin{figure*}[t] \small
	\vspace{-0.2cm}
	\centering
	\includegraphics[width=0.92\linewidth]{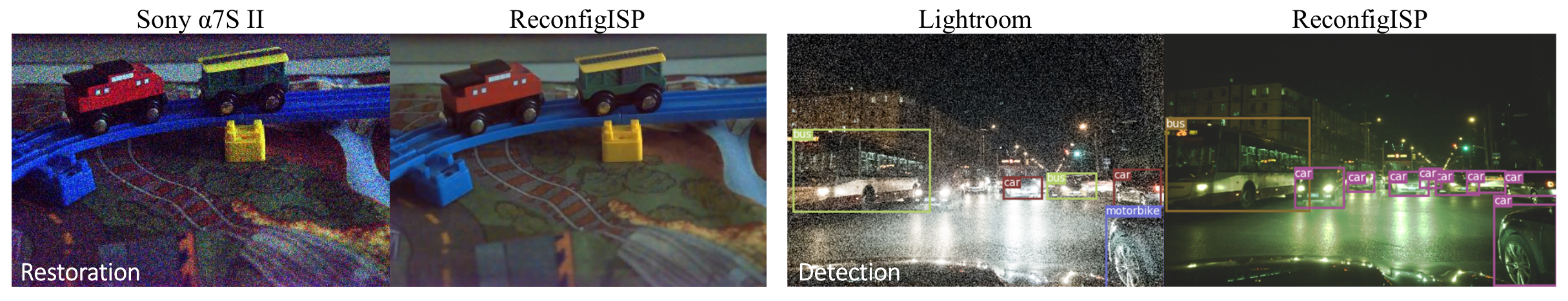}
	\vskip -0.1cm
	\caption{
		While inheriting the modularity of traditional ISP, ReconfigISP can adapt to different tasks, such as image restoration and object detection by learning to reconnect specialized ISP modules. For each task, the left image shows the results of a default camera processing pipeline while the right image presents our output (zoom in for best view). In the task of object detection, ReconfigISP optimizes the ISP for more accurate detections rather than perceptual quality.
	}
	\label{fig:introduction}
	\vspace{-0.3cm}
\end{figure*}

In this study, our goal is to keep the modular design of the traditional ISP, but learn a reconfigurable ISP architecture that can quickly adapt to various tasks (\eg, restoration and object detection), scenes (\eg, daytime and nighttime) and runtime constraints.
To this end, we present a novel and versatile \textit{reconfigurable} ISP framework, named ReconfigISP. 
The proposed framework allows changes and rerouting in the ISP architecture, \ie, what basic modules should be chosen and how they are connected, optimized towards designated objective functions. Given the flexibility, specialized modules can be combined to generate numerous ISP pipelines to handle various application scenarios, but still preserving the modularity and parameters of each module. Experimental results show that our framework achieves a significant improvement over traditional camera pipelines (Sec.~\ref{sec:experiments}). Examples are shown in Fig.~\ref{fig:introduction}.

Searching for an optimal ISP pipeline is a new and interesting problem. In this work, we make several technical contributions to address some of the non-trivial challenges:
%
\textit{(1) Tunable ISP modules} - Most existing ISP modules are non-differentiable. As a result, the parameters of these modules cannot be efficiently optimized with end-to-end training, making the search for optimal pipeline intractable.
To address this issue, we use a convolutional neural network to imitate each non-differentiable module. The proxy network observes input images and module parameters, together with global image statistics that the module may need. The network is trained to resemble the output of the original non-differentiable module. With all modules differentiable, the ISP architecture can be efficiently explored with the help of neural architecture search. 
%
\textit{(2) Proxy tuning and online pruning} - With the aforementioned framework, we still need to resolve the gap in data distribution. In particular, the data we use for ISP architecture search depend on the target task, and they might be unseen to a pre-trained proxy network. To mitigate the domain gap, we devise an effective proxy tuning approach -- fine-tuning proxy networks with the observed fresh data on the fly during ISP architecture search. Our approach updates the ISP architecture and proxy networks alternately.
And it gradually prunes modules with low performance to accelerate the training process. 

We note that there is a surge of interest in deep ISP approaches~\cite{chen2018learning,dai2020awnet,schwartz2018deepisp}, which converts an ISP pipeline to a single end-to-end deep network.
ReconfigISP offers a compelling alternative to the deep ISP paradigm. First, ReconfigISP can adjust its pipeline purely driven by objective functions, while existing deep ISP solutions do not offer such flexibility. For instance, to fulfill a specific runtime constraint, a deep ISP solution will need to redesign the network architecture from scratch.
Perhaps more notably, ReconfigISP preserves the modularity of the image formation process stages. It is interpretable as each module plays a clear role in the ISP pipeline. Hence, our method still allows human intervention if needed, and it offers insights to design principles of effective ISP pipelines beyond just a reconfigurable framework. 
Moreover, to cope with different processing functionalities, deep ISP methods typically require an extensive network with millions of parameters. Expensive retraining with a large quantity of data is needed for every new task to prevent overfitting. The data requirement is unrealistic, especially if one wishes to repurpose the ISP for complex tasks like object objection. 
In contrast, our framework only needs to tune hundreds of parameters with limited data. 
%
We will validate the above advantages of ReconfigISP in our experiments.


\section{Related Work}

\noindent\textbf{ISP Modules.}
An ISP usually involves an elaborated pipeline to handle image noise, chrominance, luminance, sharpness, \etc. 
In this study, we prepare a concise module pool, as shown in Table~\ref{tab:algorithm_pool}, with focus on common algorithms for denoising, demosaicing, tone mapping and white balancing.
More details about the modules can be found in the supplementary material.
It is noteworthy that many classic algorithms are non-differentiable, for instance, denoising algorithms like Bilateral~\cite{Tomasi1998BilateralFF}, Median~\cite{Huang1979AFT} and NLM~\cite{Buades2005ANA}, and BM3D~\cite{Dabov2006ImageDW}; and tone mapping algorithms like Reinhard~\cite{Reinhard2002PhotographicTR} and Crysisengine. We convert these algorithms to proxy networks to facilitate ISP architecture search.  Note that we also consider some lightweight deep models in our module pools, including Path-Restore~\cite{Yu2019PathRestoreLN} and DemosaicNet~\cite{gharbi2016deep}. All proxy modules will be released.

\noindent\textbf{ISP Pipelines.} 
There are several studies in the literature that focus on either traditional camera pipelines or deep ISP approaches. 
%
Hu~\etal~~\cite{hu2018exposure} propose Exposure, a white-box image post-processing framework with eight differentiable filters. 
%
Karaimer~\etal~~\cite{karaimer2016software} devise a 12-stage software ISP that allows users to study the effects of each algorithm in the context of a full ISP. However, such traditional ISP framework does not support end-to-end optimization.
Tseng~\etal~\cite{tseng2019hyperparameter} propose a proxy optimization scheme to efficiently search for the best ISP hyperparameters. Mosleh~\etal~\cite{mosleh2020hardware} further develop an evolutionary algorithm to optimize ISP parameters on the hardware.
Different from existing proxy approach~\cite{tseng2019hyperparameter}, we train a differentiable proxy for each non-differentiable algorithm instead of the whole ISP pipeline. Thus, we can explore different ISP architectures while the method of~\cite{tseng2019hyperparameter} only applies to a fixed pipeline.

Deep-learning-based ISP is drawing considerable research interest in recent years. 
Schwartz~\etal~\cite{schwartz2018deepisp} propose DeepISP that uses a deep neural network that contains a low-level stage and a high-level stage.
Dai~\etal~\cite{dai2020awnet} devise a network with attention and wavelet transform.
Chen~\etal~\cite{chen2018learning} propose SID dataset containing paired data between low light and normal illuminance. A U-Net structure trained with SID dataset significantly outperforms the traditional pipeline embedded in the camera. In this work, we use this U-Net model as a strong competitor to verify the effectiveness of our method.
The aforementioned studies adopt a fixed network to replace the ISP pipeline. The network is usually large, containing massive number of parameters that require tuning or retraining for every new task. As mentioned in the introduction, these approaches do not enjoy the same flexibility and modularity as in the proposed ReconfigISP, which is readily adaptable to different application scenarios at a low cost by adjusting both ISP architecture and parameters given the designated cost functions. 


\begin{table*}[t]\centering\small
	\caption{The module pool contains 22 algorithms. ``Domain'' represents the input and output patterns. ``Category'' describes the functionality of the algorithms. The number in the parentheses denotes the quantity of configurable parameters for each algorithm. The \textit{italic algorithms} are non-differentiable on GPU in our implementation.}
	\vskip -0.2cm
	\begin{tabular}{l|l|l}
		\hline
		\multicolumn{1}{c|}{Domain}              & \multicolumn{1}{c|}{Category} & \multicolumn{1}{|c}{Algorithm}                                                                             \\ \hline\hline
		RAW -\textgreater RAW                    & Denoising                     & \textit{Bilateral-Bayer (3)}, \textit{Median-Bayer (1)}, \textit{NLM-Bayer (3)}, Path-Restore-Bayer (0)                                                                                    \\ \hline
		RAW -\textgreater sRGB                   & Demosaicing                   & \textit{Laplacian (0)} \cite{Wu2004ColorIF}, Nearest (0), Bilinear (0) \cite{Getreuer2011LinearMF}, DemosaicNet (0) \cite{gharbi2016deep}      \\ \hline
		\multirow{4}{*}{sRGB -\textgreater sRGB} & Denoising                     &
		\textit{Bilateral (3)} \cite{Tomasi1998BilateralFF}, \textit{Median (1)} \cite{Huang1979AFT}, \textit{NLM (3)} \cite{Buades2005ANA}, \textit{BM3D (5)} \cite{Dabov2006ImageDW}, Path-Restore (0) \cite{Yu2019PathRestoreLN} \\ \cline{2-3}
		& Gamma Correction              & Gamma (1)                                                                                                 \\ \cline{2-3}
		& Global Tone Mapping           &  \textit{Reinhard (2)} \cite{Reinhard2002PhotographicTR}, \textit{Crysisengine (1)}, \textit{Filmic (2)} \cite{Eilertsen2017ACR}, Manual (3) \\ \cline{2-3}
		& White Balance                 & \textit{Whitepatch (1)} \cite{Rizzi2002ColorCB}, Grayworld (0), Linear (3), Quadratic (30) \cite{schwartz2018deepisp}      \\ \hline
	\end{tabular}
	\label{tab:algorithm_pool}
	\vspace{-0.2cm}
\end{table*}



\section{Methodology}

The overview framework of ReconfigISP is shown in Fig.~\ref{fig:framework}. In Sec.~\ref{subsec:isp_proxy}, we introduce differentiable proxy networks to replace non-differentiable modules. In Sec.~\ref{subsec:isp_nas}, the algorithm for ISP architecture search is discussed. Finally, in Sec.~\ref{subsec:isp_param_opt}, we explain how module parameters are fine-tuned after the ISP architecture is fixed.

\subsection{Differentiable Proxy Networks}
\label{subsec:isp_proxy}

Given the module pool shown in Table~\ref{tab:algorithm_pool}, our goal is to search for the most effective ISP pipeline composed of these algorithms. Some of the algorithms are intrinsically differentiable and thus can be optimized in an end-to-end manner, while others are non-differentiable, causing difficulties in architecture and parameter optimization. To address this challenge, we make a differentiable proxy network for each non-differentiable module.

\begin{figure*}[t] \small
	\vspace{-0.2cm}
	\centering
	\includegraphics[width=0.88\linewidth]{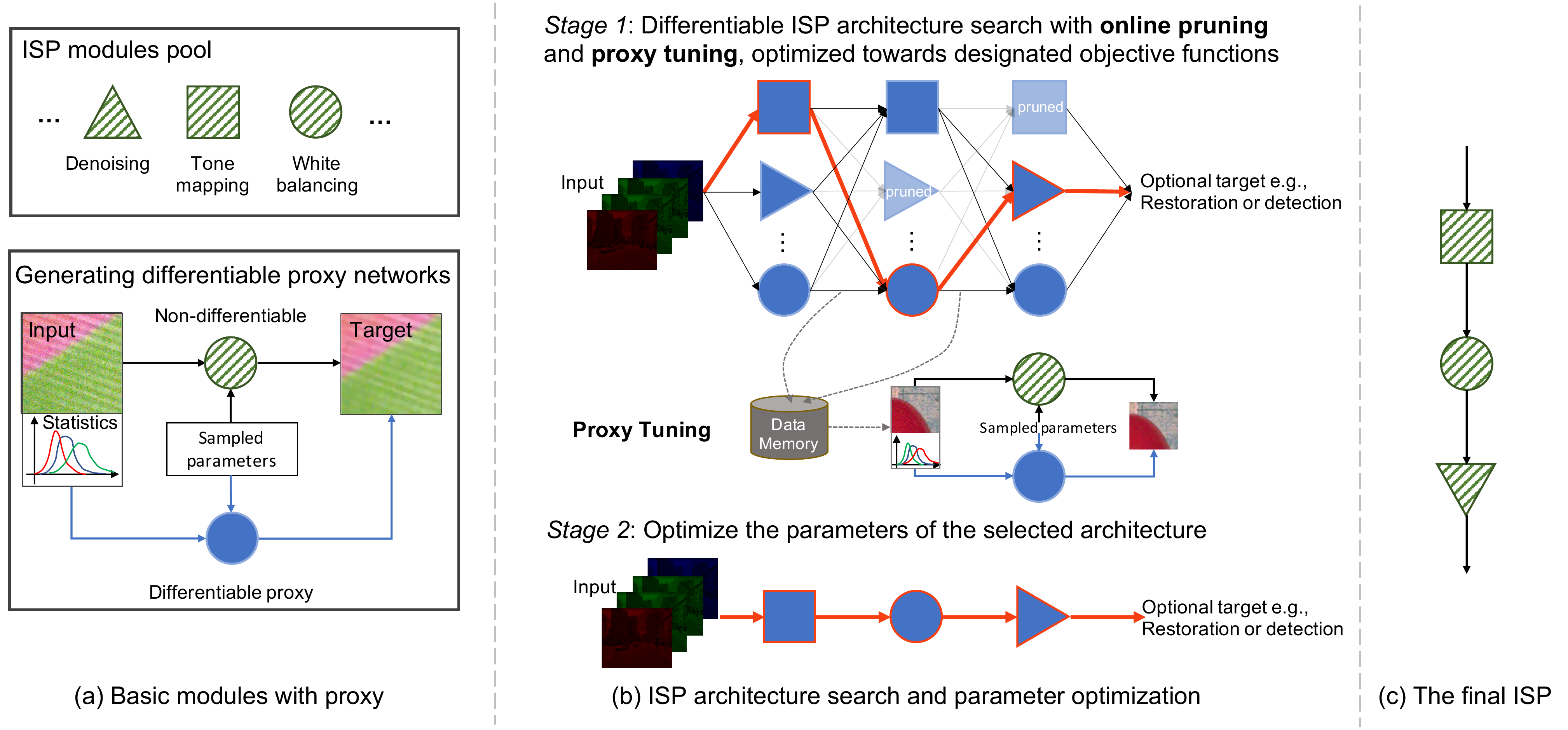}
	\vskip -0.3cm
	\caption{
		The overview of ReconfigISP. (a) We implement a comprehensive set of ISP modules ($N=22$), and construct a differentiable proxy for each non-differentiable algorithm. (b) In stage 1, we search for the optimal ISP architecture using Darts~\cite{liu2019darts} with the newly proposed online pruning and proxy tuning. Finally, in stage 2, the algorithm parameters of the selected architecture are fine-tuned for the best performance. (c) The final ISP adopts the searched architecture with original ISP modules and optimized parameters.
	}
	\label{fig:framework}
	\vspace{-0.2cm}
\end{figure*}

Formally, let $\hat{f}_j$ denote the $j$-th module and $f_j$ denote its proxy network with weights $w_j$. Suppose $\hat{f}_j$ is differentiable, and we need not apply any change to this algorithm, \ie, $f_j = \hat{f}_j$. 
%
As shown in Fig.~\ref{fig:framework}~(a), to guarantee that the proxy network resembles the original algorithm for different input images $\mathbf{X}$ and parameters $p_j$, we randomly sample images and parameters while training the proxy network. Global image statistics are extracted for non-local modules. Let $L_p$ denotes the fidelity loss for proxy training, \eg, L1 or L2 loss. The objective function is then formulated as $L_p\left(f_j(\mathbf{X}, p_j; w_j), \hat{f}_j(\mathbf{X}, p_j)\right)$.

\subsection{Reconfigurable ISP Architecture}
\label{subsec:isp_nas}

\noindent\textbf{Differentiable ISP Architecture Search.}
Our ISP architecture search is inspired by existing work on neural architecture search (NAS). Considering that NAS based on reinforcement learning~\cite{Baker2017,Zoph2017} and evolutionary methods~\cite{real2019regularized} is expensive, we adopt the more efficient differentiable NAS. In particular, we choose Darts~\cite{liu2019darts}, with our newly proposed online pruning and proxy tuning to solve the specific challenges in ISP. Details are provided as follows.

Given all the differentiable proxies at hand, we incorporate these small ISP modules into a super network for architecture search, as shown in Fig.~\ref{fig:framework}~(b). Let $K$ denotes the maximum length of the desired ISP pipeline. At each step $k$, there are $N+1$ modules parallelly connected, \ie, $N$ differentiable algorithms and one skip connection module.
Note that not all modules are valid at each step. For example, an algorithm designed for sRGB images cannot work in the RAW domain. Therefore, we need to specify the valid domain for each step, and then a set of valid index $V$ can be derived. If $(k,j)\in{V}$, then the $j$-th module is valid at step $k$.
For every valid module $f_{kj}$, it is assigned an architecture weight $\alpha_{kj}$, representing how likely it is to select this module. All the architecture weights should be non-negative and sum to one within a single step, \ie, $\sum_{j=1}^{N}\alpha_{kj}=1, \text{s.t.} ,\alpha_{kj}>0, (k,j) \in V$. 

Next, we briefly introduce the inference process of the constructed super network. Let $\mathbf{X}$ denote the input image. The intermediate input image at step $k$ is denoted by $\mathbf{X}_{k-1}$. The network inference at step $k$ can be formulated as:
\vspace{-0.2cm}
\begin{equation}
\label{eq:inference}
\mathbf{X}_k = \sum\nolimits_{j=1}^{N} \alpha_{kj} f_{j}(\mathbf{X}_{k-1}, p_{kj}; w_{j}), \ \text{s.t.}\ (k,j) \in V,
\end{equation}
%
where $p_{kj}$ represents the input parameters of the algorithm $f_{j}$ at step $k$, and $w_{j}$ denotes the weight of proxy network if $f_{j}$ is a differentiable proxy. For example, suppose $f_{j}$ denotes the proxy network of a median filter, then $p_{kj}$ is the filter size at $k$-th step, and $w_{j}$ represents the network weights.

We alternately update the architecture $\alpha_{kj}$ and the parameters of each ISP module $p_{kj}$. Let $L(\hat{\mathbf{X}}, \mathbf{Y})$ denote the loss function, where $\hat{\mathbf{X}} = \mathbf{X}_K$ is the output of the ISP pipeline, and $\mathbf{Y}$ is the target image or label. The loss function depends on the specific task. To update the algorithm parameters, we derive the gradients as:
\begin{equation}
\label{eq:parameter_step}
\delta_{{p}_{kj}} = \nabla_{p_{kj}}L(\hat{\mathbf{X}}, \mathbf{Y}; p, \alpha),
\end{equation}
where $p$ and $\alpha$ denote all the algorithm parameters $\{p_{kj}\}$ and architecture weights $\{\alpha_{kj}\}$, respectively.

As for the update of architecture, directly applying the gradients is suboptimal. As discussed in Darts~\cite{liu2019darts}, the optimal algorithm parameters may vary according to the specific architecture. Thus, we should also consider the algorithm parameters while optimizing the architecture. Specifically, we conduct one step of meta learning to get better architecture gradients:
\begin{equation}
\label{eq:architecture_step}
\begin{aligned}
	\hat{p} &= p - \xi\nabla_{p}L(\hat{\mathbf{X}}_{train}, \mathbf{Y}_{train}; p, \alpha), \\
	\delta_{{\alpha}_{kj}} &= \nabla_{\alpha_{kj}}L\left(\hat{\mathbf{X}}_{val}, \mathbf{Y}_{val}; \hat{p}, \alpha\right), 
\end{aligned}
\end{equation}
where $\xi$ represents the learning rate for meta training. The data are divided into two groups, for meta training and meta validation, respectively.

\noindent\textbf{Online Pruning.}
The aforementioned searching algorithm is still time consuming in practice due to the expensive structure of the super network. Each algorithm is repeatedly conducted at every step, resulting in a huge quantity of computations. To alleviate this computational burden, we prune those candidate modules with low architecture weights.

An intuitive pruning strategy is to set a hard threshold of the architecture weight. The module whose architecture weight is smaller than the threshold should be discarded. However, we found such strategy sensitive to the optimization hyper-parameters. The pruning results can be different with multiple trials. Hence, we set a relative threshold according to the architecture weight with the highest potential. In particular, the pruning mechanism can be formulated as
\begin{equation}
\label{eq:pruning}
\left\{
\begin{array}{ll}
(k, j) \in V,    & \quad\alpha_{kj} > \eta\max_{j}\alpha_{kj},\\
(k, j) \notin V, & \quad{\text{otherwise}},
\end{array}
\right.
\end{equation}
where $\eta \in (0, 1)$ is the relative threshold.

\noindent\textbf{Proxy Tuning.}
As we may apply ReconfigISP to different tasks, the data distribution can be very diverse. Thus, the proxy training data may not cover the data distribution of the target task. Moreover, the intermediate result $\mathbf{X}_k$ is a combination of the outputs from multiple algorithms, which is unseen in the proxy training process. Therefore, the proxy network may fail to approach the original algorithm given this data distribution gap. To mitigate this issue, we propose proxy tuning, which fine-tunes the proxy networks using the observed new data during the architecture search.

The process of proxy tuning is illustrated in Fig.~\ref{fig:framework}~(b). Specifically, we build a data memory $M$ to store the intermediate outputs from the super network. The data memory is organized as a queue, with a maximum capacity of $|M|_\text{max}$. At each training timestep $t$, the intermediate results $\mathbf{X}_1,\dots,\mathbf{X}_K$ are added to the queue, and some past images are removed if the queue is full.
At every $t_p$ step, a batch of data $\mathbf{X}_m$ are randomly sampled from the data memory and backpropagation is applied to derive the gradients of proxy network weights:
\begin{equation}
\label{eq:proxy_tuning}
\delta_{w_{j}} = \nabla_{w_{j}}L_p\left(f_{j}(\mathbf{X}_m, \tilde{p}_{j}; w_{j}), \hat{f}_j(\mathbf{X}_m, \tilde{p}_{j})\right),
\end{equation}
where $\hat{f}_j$ denotes the original algorithm corresponding to the proxy network $f_{j}$, and $\tilde{p}_{j}$ represents a set of randomly sampled parameters for this algorithm. $L_p$ denotes the loss function for proxy training. We do not perform proxy tuning on the module that does not need a proxy network.

\noindent\textbf{Summary.}
The overall procedure for ISP architecture search is illustrated in Fig.~\ref{fig:framework}~(b). The searching algorithm is summarized in Algorithm~\ref{alg:ispas}. At each timestep $t$, we update architecture and algorithm parameters, refresh the data memory, and perform online pruning to cut off unnecessary modules. At every $t_p$ steps, we conduct proxy tuning to ensure the effectiveness of each proxy network.

\begin{algorithm}[t]
	\caption{ISP Architecture Search}
	\label{alg:ispas}
	\small
	\begin{algorithmic}
		\State Prepare for the module pool $\{f_{j}\}$ with proxy weights $\{w_j\}$
		\State Initialize algorithm parameters $\{p_{kj}\}$ and architecture $\{\alpha_{kj}\}$
		\State Initialize data memory $M=\emptyset$, set the memory size $|M|_\text{max}$
		\State Specify learning rate $\gamma$, total iterations $T$, tuning interval $t_p$
		\For{$t=1,T$}
		\State Sample training data $\mathbf{X}_{train}$, $\mathbf{Y}_{train}$, $\mathbf{X}_{val}$, $\mathbf{Y}_{val}$
		\State $\alpha_{kj} \gets \alpha_{kj} - \gamma\delta_{{\alpha}_{kj}}$\Comment{Update architecture, Eq.~(\ref{eq:architecture_step})}
		\State $p_{kj} \gets p_{kj} - \gamma\delta{{p_{kj}}}$\Comment{Update parameters, Eq.~(\ref{eq:parameter_step})}
		\State Update memory $M$ using intermediate data $\mathbf{X}_{k}$
		\State Update valid module $V=\{(k,j)\}$\Comment{Pruning, Eq.~(\ref{eq:pruning})}
		\If{$t \equiv 0 \mod t_p$}
		\State Sample data $\mathbf{X}_m$ from memory $M$
		\State $w_{j} \gets w_{j} - \gamma\delta_{w_{j}}$\Comment{Proxy tuning, Eq.~(\ref{eq:proxy_tuning})}
		\EndIf
		
		\EndFor
	\end{algorithmic}
\end{algorithm}

\subsection{ISP Parameters Optimization}
\label{subsec:isp_param_opt}
After the architecture search, we select the ISP pipeline with the highest architecture weights. Proxy networks are replaced back with the original modules (Fig.~\ref{fig:framework}~(c)). In particular, at step $k$, the index of the selected module is $a_k=\argmax_{j}\alpha_{kj}$, and the selected module is denoted by $f_{a_k}$. Given the searched ISP architecture, the algorithm parameters are not optimal, because there are still multiple algorithms not pruned at each step in the super network. Therefore, we further optimize the algorithm parameters specialized for the selected ISP pipeline. The inference process at the $k$-th step is formulated as $\mathbf{X}_k = f_{a_k}(\mathbf{X}_{k-1},p_{a_k};w_{a_k})$, where $p_{a_k}$ is the algorithm parameters we aim to optimize, and $w_{a_k}$ stands for the network weights after proxy tuning.

The loss function depends on the specific task. For image restoration, we adopt L2 loss, $L=||\hat{\mathbf{X}} - \mathbf{Y}||_2^2$, where $\hat{\mathbf{X}}=\mathbf{X}_K$ and $\mathbf{Y}$ denote the output image and the ground truth, respectively. As for object detection, we follow the design of YOLOv3~\cite{redmon2018yolov3}. The loss function is composed of a localization term and a classification term to penalize coordinate errors and class label mismatch, respectively.
Besides the above loss functions, ReconfigISP can be optimized for any objective function as long as backpropagation is applicable.

\section{Experiments}
\label{sec:experiments}

To facilitate quantitative analysis, we set the total length of ISP as $K=5$,\footnote{Note that this is a simplified pipeline since in practice an ISP pipeline contains many modules. But our method still shows competitive results in comparison to the very deep ISP approach (see Sec.~\ref{subsec:compare_deepisp}).} with one step in the RAW domain, one demosaicing step, and three steps in the sRGB domain. 
In this specific setting, the overall ReconfigISP framework contains 226 architecture and module parameters in total.
We adopt L2 loss for proxy training. The data memory size is chosen as 1,000. The online tuning threshold $\eta$ is set as 0.2. The proxy tuning interval $t_p$ is 20.

The total number of iterations are 200,000 and 80,000 for architecture search and parameter optimization, respectively. The learning rate $\gamma$ and meta learning rate $\xi$ are the same. The initial value is $1\times10^{-4}$, and then it is decayed by a half every quarter of the training process. SRCNN~\cite{dong2016image} is used as the proxy network, and the differentiable proxies are pre-trained on SIDD~\cite{SIDD_2018_CVPR} dataset.
We use Adam~\cite{kingma2015adam} optimizer and implement our framework on PyTorch~\cite{paszke2017automatic}. Our experiments are conducted on four NVIDIA GeForce GTX 1080 GPUs.

\subsection{Evaluation on Image Restoration}
To validate the effectiveness of the proposed method for different sensors, scenes and light conditions, we select two challenging benchmarks that the proxy networks have never observed -- SID dataset~\cite{chen2018learning} containing images captured by a DSLR camera, and S7 ISP dataset~\cite{schwartz2018deepisp} that is composed of smartphone images.


\begin{table}[t]\centering\small
		\caption{Quantitative results on SID Dataset~\cite{chen2018learning}.}
		\vskip -0.2cm
		\setlength{\tabcolsep}{3mm}
		\begin{tabular}{c|c|c|c|c}
			\hline
			\multirow{2}{*}{\begin{tabular}[c]{@{}c@{}}SID\\ 0.1s to 10s\end{tabular}} & \multicolumn{2}{c|}{RGB} & \multicolumn{2}{c}{Gray} \\ \cline{2-5}
			& PSNR       & SSIM        & PSNR        & SSIM        \\ \hline\hline
			Sony $\alpha$7 II            & 15.69      & 0.1550      & 20.34       & 0.3352      \\			
			ReconfigISP            & \textbf{25.65}      & \textbf{0.7527}      & \textbf{31.90}       & \textbf{0.9008}      \\ \hline
		\end{tabular}
		\label{tab:sid}
\end{table}

\begin{table}[t]\centering\small
		\caption{Quantitative results on S7 ISP Dataset~\cite{schwartz2018deepisp}.}
		\vskip -0.2cm
		\setlength{\tabcolsep}{3mm}
		\begin{tabular}{c|c|c|c|c}
			\hline
			\multirow{2}{*}{S7 ISP} & \multicolumn{2}{c|}{RGB} & \multicolumn{2}{c}{Gray} \\ \cline{2-5}
			& PSNR       & SSIM        & PSNR        & SSIM        \\ \hline\hline
			Samsung S7                  & 21.08      & 0.4518      & 23.99       & 0.6297      \\ 
			ReconfigISP                    & \textbf{23.31}      & \textbf{0.7007}      & \textbf{26.83}       & \textbf{0.7697}      \\ \hline
		\end{tabular}
		\label{tab:s7isp}
\end{table}

\begin{figure}[t] \small
	\centering
	\includegraphics[width=0.95\linewidth]{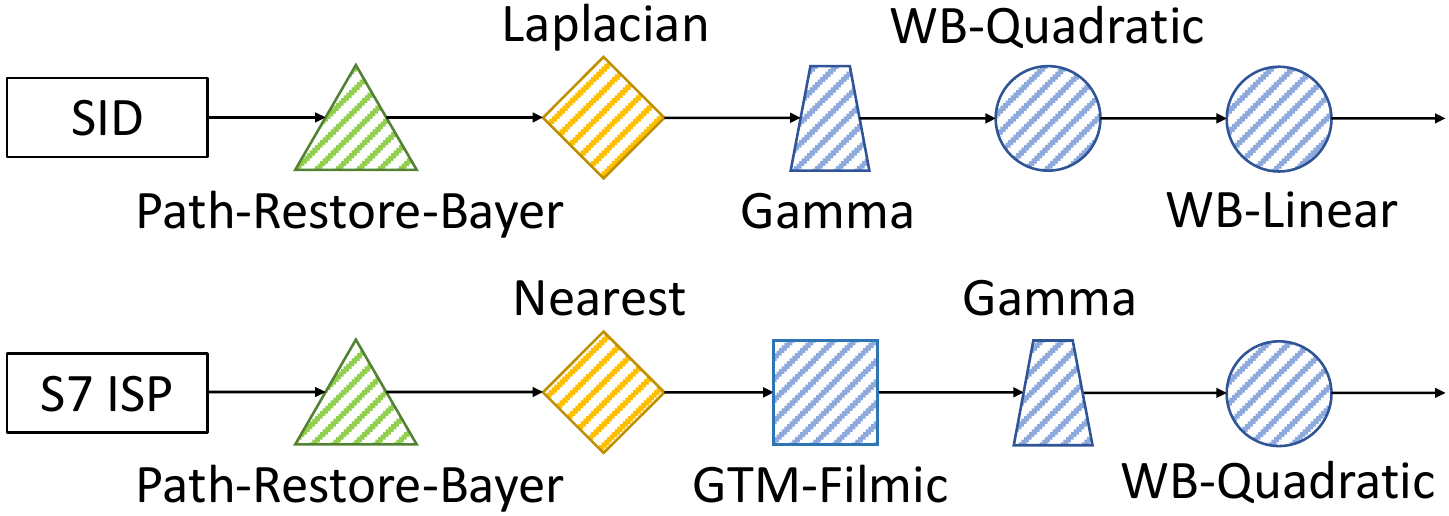}
	\vskip -0.2cm
	\caption{
		The searched ISP architecture on SID and S7 ISP datasets for image restoration. The green, yellow and blue colors represent the mapping domain of RAW$\rightarrow$RAW, RAW$\rightarrow$sRGB and sRGB$\rightarrow$sRGB, respectively. GTM and WB stand for global tone mapping and white balance, respectively.
	}
	\label{fig:isp_sid_s7}
	\vspace{-0.2cm}
\end{figure}

\begin{figure}[t] \small
	\vspace{-0.2cm}
	\centering
	\includegraphics[width=0.88\linewidth]{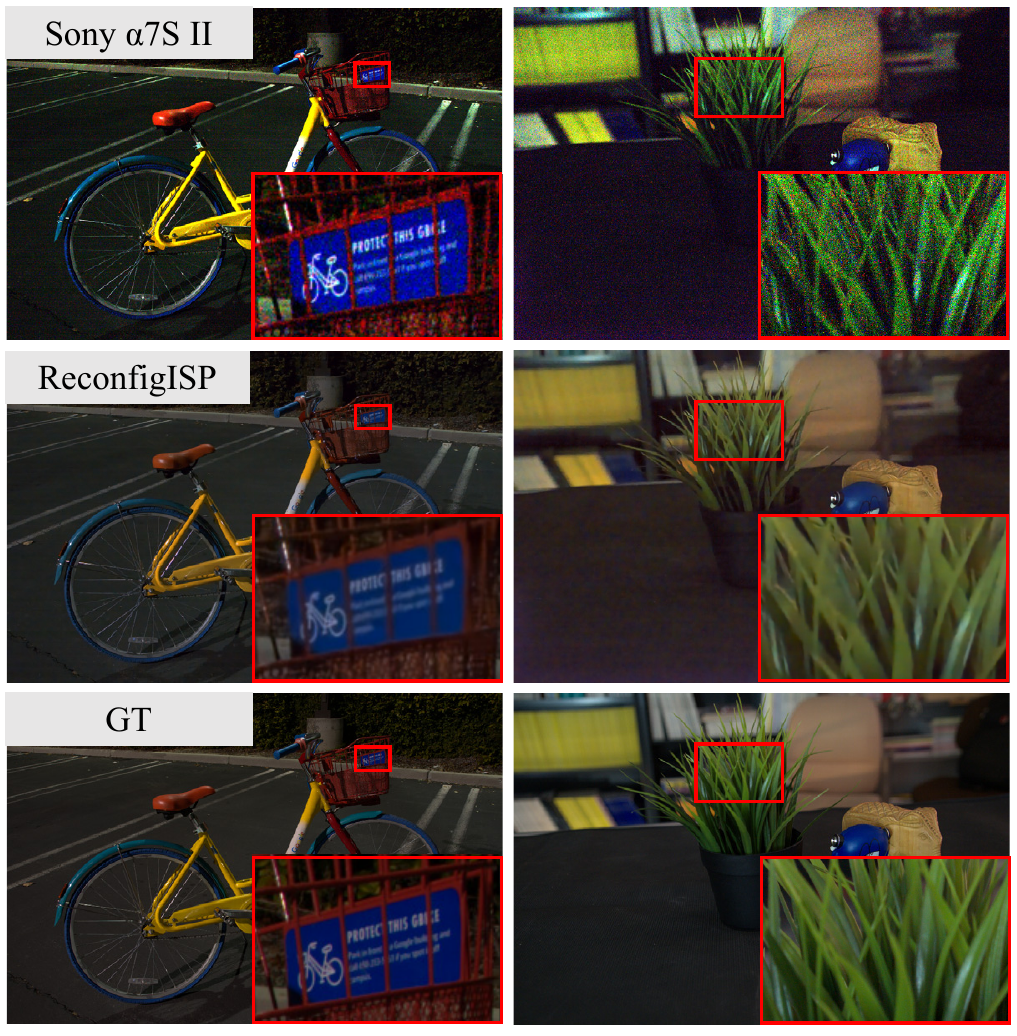}
	\vskip -0.2cm
	\caption{
		Qualitative results on SID dataset~\cite{chen2018learning}. The input and ground-truth exposure time are 0.1s and 10s, respectively.
	}
	\label{fig:exp_sid}
	\vspace{-0.2cm}
\end{figure}

\begin{figure}[t] \small
	\centering
	\includegraphics[width=0.88\linewidth]{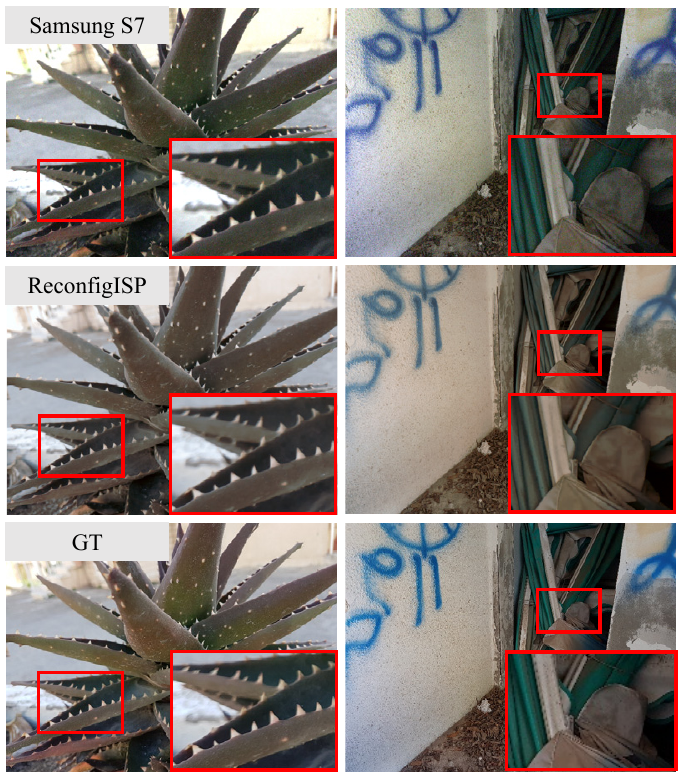}
	\vskip -0.2cm
	\caption{
		Qualitative results on S7 ISP dataset~\cite{schwartz2018deepisp}.
	}
	\label{fig:exp_s7isp}
	\vspace{-0.35cm}
\end{figure}

\noindent\textbf{SID Dataset.}
We verify the effectiveness of our method on SID dataset~\cite{chen2018learning}, which contains several challenging images captured under extreme low light.
We conduct experiments on the Sony subset, where images are captured by Sony $\alpha$7S II.
To verify the superiority of our algorithm on small-scale data, we randomly select 5\% of the training images. Without loss of generality, we consider a specific setting with 0.1s short exposure and 10s long exposure.

The searched ISP pipeline is shown in Fig.~\ref{fig:isp_sid_s7} (top). A deep denoising method Path-Restore-Bayer is chosen in the RAW domain. The Laplacian interpolation is used for demosaicing. In the sRGB domain, gamma correction is applied followed by two white balancing modules, indicating that it is challenging to adjust the white balance in extreme low-light conditions.

Quantitative comparisons are shown in Table~\ref{tab:sid}. Sony $\alpha$7 II represents the results produced by the default camera pipeline, multiplied by an exposure compensation coefficient of 100 (10s divided by 0.1s). We present the PSNR and SSIM results of color and gray images in Table~\ref{tab:sid}. 
Our ReconfigISP surpasses the default camera ISP by a large margin, \ie, more than 10 dB improvement.
Qualitative results are presented in Fig.~\ref{fig:exp_sid}. Compared to the Sony pipeline, our method yields cleaner results with less noise, and the exposure looks more natural. It is worth noting that our method achieves such compelling performance with only hundreds of parameters being tuned.

\noindent\textbf{S7 ISP Dataset.}
S7 ISP~\cite{schwartz2018deepisp} is a dataset collected from Samsung S7 rear camera. Following the setting of SID, we only use 5\% of the training data. The quantity of validation and test subsets is not changed.

The searched ISP architecture is shown in Fig.~\ref{fig:isp_sid_s7} (bottom). The pipeline is slightly different from that of SID. In particular, Nearest interpolation is used instead of Laplacian filter. Moreover, a global tone mapping method Filmic is applied prior to gamma correction. We observe that Filmic works well for S7 ISP dataset, but it may amplify artifacts when applied to SID dataset, where the noise is severe under extreme low light.

Quantitative results are presented in Table~\ref{tab:s7isp}. 
It is observed that ReconfigISP is superior to the Samsung camera pipeline by more than 2 dB on both RBG and gray images. 
Visual results are shown in Fig.~\ref{fig:exp_s7isp}. The default Samsung ISP fails to remove the color noise in low-light condition. On the contrary, ReconfigISP yields clean outputs while preserving the details. 

\subsection{Adapting to Specific Efficiency Constraints}
\label{subsec:efficiency}
In real-world applications, we may have different efficiency constraints because of limited computational resources and the demand for real-time display. ReconfigISP adjusts the performance-complexity trade-off by introducing an efficiency term in the loss function. Specifically, during ISP architecture search, the loss $L$ is multiplied by ${(Lat)}^\beta$, where $Lat$ stands for the latency of the current ISP. The larger $\beta$ is, the more efficient pipeline will ReconfigISP select. In particular, ReconfigISP, ReconfigISP-Fast and ReconfigISP-Faster correspond to $\beta=0$, $\beta=0.14$ and $\beta=0.28$, respectively. Note that such a flexibility is not available in traditional ISP pipelines and deep ISP approaches.

The performance and CPU\footnote{Intel(R) Core(TM) i7-8700 CPU @ 3.20GHz} runtime of different ISP frameworks are presented in Table~\ref{tab:time}. Camera ISP represents the camera processing pipeline of the corresponding dataset. ReconfigISP-Faster achieves up to 50 times speedup compared to ReconfigISP, at the expense of performance. We observe that the acceleration is attributed to fewer computations for denoising. Nevertheless, ReconfigISP-Faster is still better than or comparable to the default camera ISP. The selected ISP architectures are described in the supplementary material. 

\begin{table}[t]\centering\small
	\caption{Results of different efficiency constraints. We report CPU runtime to process one megapixel.}
	\vskip -0.2cm
	\begin{tabular}{c|c|c|c|c}
		\hline
		\multirow{2}{*}{Dataset and Metric} & \multicolumn{2}{c|}{SID~\cite{chen2018learning}} & \multicolumn{2}{c}{S7 ISP~\cite{schwartz2018deepisp}} \\ \cline{2-5}
		& PSNR       & Time (s)        & PSNR        & Time (s)       \\ \hline\hline
		Camera ISP					 & 15.69      & -          & 21.08       & - \\
		ReconfigISP                  & 25.65      & 1.16       & 23.31       & 1.53      \\ 
		ReconfigISP-Fast             & 23.72      & 0.63       & 22.70       & 0.61      \\
		ReconfigISP-Faster           & 20.55      & 0.049      & 20.42       & 0.031      \\ \hline
	\end{tabular}
	\label{tab:time}
\end{table}

\begin{figure*}[t] \small
	\vspace{-0.2cm}
	\centering
	\includegraphics[width=0.92\linewidth]{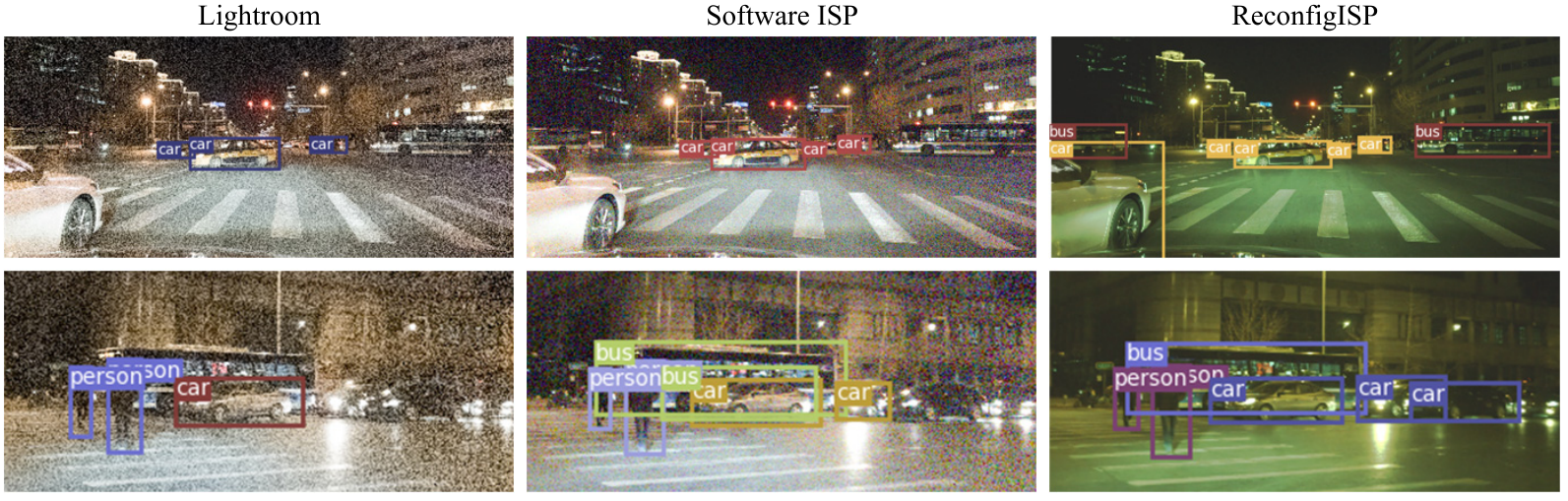}
	\vskip -0.2cm
	\caption{
		Qualitative results on OnePlus dataset for object detection (Zoom in for best view).
	}
	\label{fig:oneplus}
	\vspace{-0.3cm}
\end{figure*}

\begin{figure}[t] \small
	\vspace{-0.1cm}
	\centering
	\includegraphics[width=0.95\linewidth]{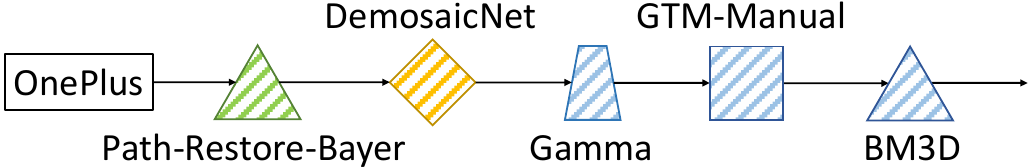}
	\vskip -0.2cm
	\caption{
		The searched ISP architecture for object detection.
	}
	\label{fig:isp_oneplus}
	\vspace{-0.1cm}
\end{figure}

\subsection{Evaluation on Object Detection}
To verify the effectiveness of the proposed framework on high-level tasks such as object detection, we collect a dataset, named OnePlus,\footnote{The dataset will be available on the project page.} containing several driving scenes in low-light condition. We use OnePlus 6T A6010 smartphone to take photos, and the sensor type is Sony Exmor IMX519. To achieve low-light constraint, we set ISO$=$6,400, exposure time$=$0.01s. With the help of Lightroom, we simultaneously capture RAW and JPEG images. The JPEG images are produced by the default ISP pipeline of Lightroom.

We focus on three classes of object in the street scenes -- person, car and bus. We use LabelMe~\cite{barriuso2012notes} to annotate 141 images, with 50 images for training and the remaining 91 images for evaluation. In the experiments, we adopt YOLOv3~\cite{redmon2018yolov3} as the detector. For inference, YOLOv3 observes the images generated by ReconfigISP and predicts the objects. As for backpropagation, YOLOv3 passes the gradients back to ReconfigISP to facilitate architecture search. Note that YOLOv3 is not updated in the training process.

The selected ISP architecture is presented in Fig.~\ref{fig:isp_oneplus}. We observe that this pipeline is significantly different from those for image restoration (see Fig.~\ref{fig:isp_sid_s7}). In particular, a deep model DemosaicNet~\cite{gharbi2016deep} is used for demosaicing. Denoising is not only performed in the RAW domain, but also conducted in the sRGB domain using BM3D~\cite{Dabov2006ImageDW}. A global tone mapping method is chosen to enlarge the image contrast. Interestingly, white balance is not adjusted in the whole pipeline, indicating that the color may be less important than noise and contrast, for detection task.

Quantitative comparisons are shown in Table~\ref{tab:oneplus}. We report the Average Precision (AP) of each class together with the mean AP (mAP). ReconfigISP is compared with Lightroom and Software ISP~\cite{karaimer2016software}. ReconfigISP surpasses the default Lightroom pipeline by a large margin. Compared with Software ISP~\cite{karaimer2016software}, ReconfigISP achieves better performance on all three classes.
We present qualitative results in Fig.~\ref{fig:oneplus}. ReconfigISP is significantly superior to the other two methods, giving accurate object classes and locations under low-light condition. Such optimized performance is achieved despite the discoloration, recapitulating that color is not as important in object detection task.

\subsection{Further Analysis}

\begin{table}[t] \small
	\caption{Object detection results on OnePlus Dataset.}
	\vskip -0.2cm
	\centering
	\setlength{\tabcolsep}{3mm}
	\begin{tabular}{c|c|c|c|c}
		\hline
		OnePlus           & Person & Car & Bus & mAP \\ \hline\hline
		Lightroom         & 0.318     & 0.469  & 0.167  & 0.318 \\ 
		Software ISP~\cite{karaimer2016software}      & 0.427     & 0.659  & 0.458  & 0.515 \\ 
		ReconfigISP             & \textbf{0.515}     & \textbf{0.697}  & \textbf{0.592}  & \textbf{0.601} \\ \hline
	\end{tabular}
	\label{tab:oneplus}
	\vspace{-0.3cm}
\end{table}


\noindent\textbf{The Effects of ISP Architecture.}
To verify the importance of ISP architecture search, we compare our method to a random search baseline. In particular, the ISP architecture is randomly sampled and tuned. The best ISP is chosen among more than 1,000 optimized pipelines. This gives the random baseline an unfair advantage. 
To have a complete analysis, we swap the ISP architectures optimized for different tasks and observe the performance change. This analysis is conducted on S7 ISP dataset for image restoration and OnePlus dataset for object detection. The corresponding ISP architectures are denoted by ReconfigISP-S7 (Fig.~\ref{fig:isp_sid_s7}) and ReconfigISP-OnePlus (Fig.~\ref{fig:isp_oneplus}), respectively. After exchanging the ISP architecture, the algorithm parameters are further fine-tuned. 

The results are presented in Table~\ref{tab:ablation}. On S7 ISP dataset, the optimal architecture surpasses ReconfigISP-OnePlus by nearly 3 dB. On OnePlus detection dataset, the specialized pipeline is superior to the S7 architecture by a large margin. Moreover, the specialized ReconfigISP consistently achieves better performance compared to random search. These results demonstrate that architecture optimization is crucial for ISP, and our ReconfigISP successfully identifies appropriate ISP architectures for diverse target tasks.

\noindent\textbf{The Effects of Proxy Tuning.}
We investigate the importance of proxy tuning on S7 ISP dataset. Without proxy tuning, the last module ``WB-Quadratic'' is replaced with ``WB-Grayworld'', which is a simple white balancing algorithm without any tuned parameters. After optimizing the algorithm parameters for this ISP architecture, the PSNR and SSIM performance are 21.01 dB and 0.6811, respectively. Compared to ReconfigISP-S7 with PSNR 23.31 dB and SSIM 0.7007, the new architecture without proxy tuning suffers a significant performance drop. We attribute this phenomenon to the data distribution gap between different tasks and datasets. When a proxy network fails to imitate the original algorithm, the architecture search becomes less effective. Thus, proxy tuning mechanism is essential to
our ReconfigISP framework.

\subsection{Comparisons to Deep ISP}
\label{subsec:compare_deepisp}

\begin{table}[t]\centering\small
	\caption{Ablation study on the ISP architecture.}
	\vskip -0.2cm
	\setlength{\tabcolsep}{3mm}
	\begin{tabular}{c|c|c|c}
		
		\hline
		
		Dataset & \multicolumn{2}{c|}{S7 ISP~\cite{schwartz2018deepisp}} & OnePlus      \\ \hline
		
		Metric        &PSNR        &SSIM        & mAP \\ \hline\hline
		
		Random Search       & 23.13        & 0.6897       & 0.566 \\
		
		ReconfigISP-S7      & \textbf{23.31}        & \textbf{0.7007}       & 0.352\\ 
		
		ReconfigISP-OnePlus & 20.55        & 0.6846       & \textbf{0.601}\\ \hline
		
	\end{tabular}
	
	\label{tab:ablation}
\end{table}

\begin{table}[t]\centering\small
	\caption{Results with different data quantity on SID~\cite{chen2018learning} dataset.}
	\vskip -0.2cm
	\setlength{\tabcolsep}{2.5mm}
	\begin{tabular}{c|c|c|c}
		\hline
		Number of Training Patches               & 100 & 500 & 3,000 \\ \hline\hline
		U-Net~\cite{chen2018learning}     & 18.43 & 24.13 & \textbf{26.62} \\
		ReconfigISP                       & \textbf{22.73} & \textbf{24.98} & 25.61 \\ \hline
	\end{tabular}
	\label{tab:train_data}
	\vspace{-0.2cm}
\end{table}

\begin{figure}[t] \small
	\vspace{-0.2cm}
	\centering
	\includegraphics[width=0.88\linewidth]{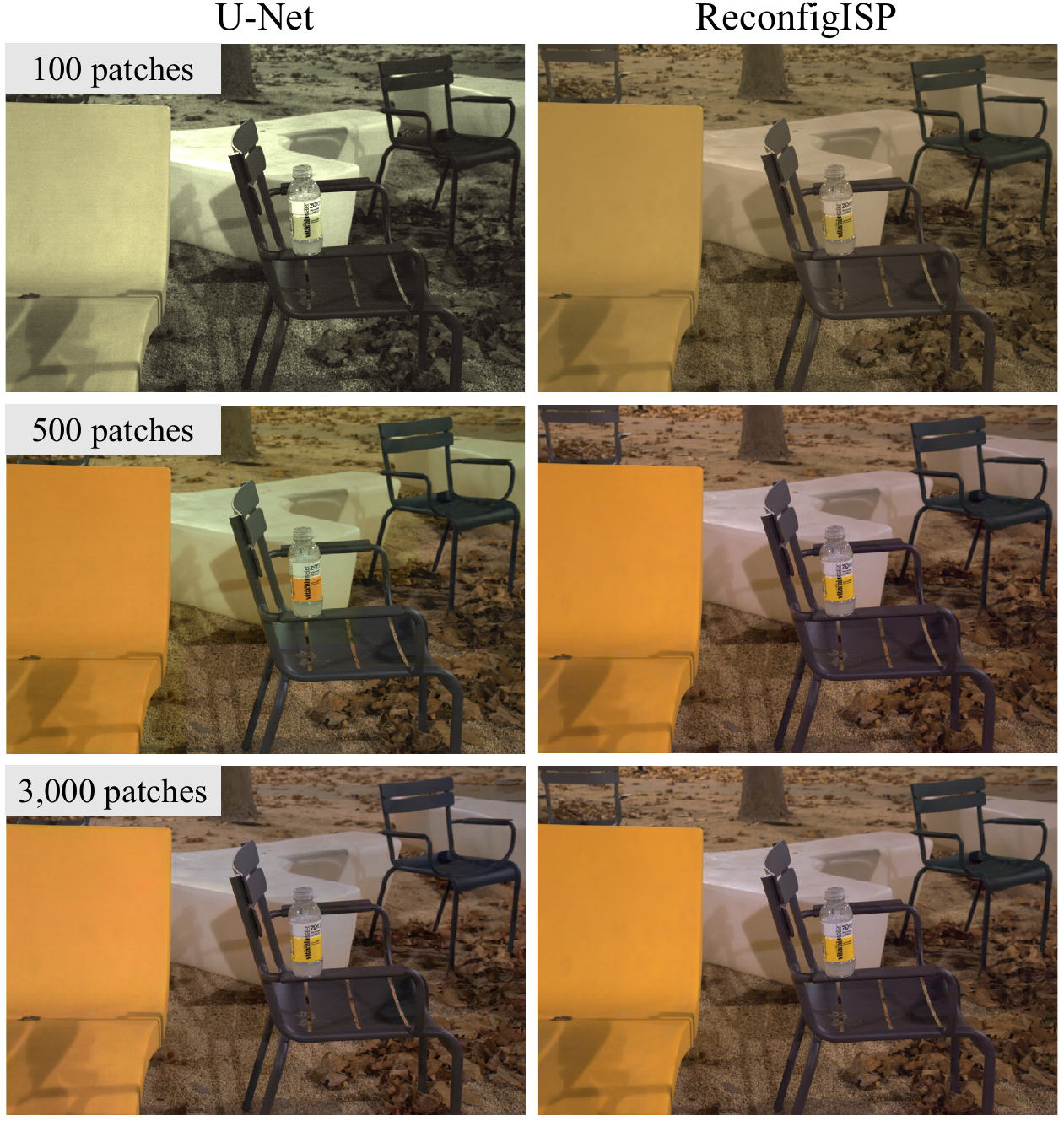}
	\vskip -0.2cm
	\caption{
		Qualitative comparisons to U-Net~\cite{chen2018learning} with different quantity of training data.
	}
	\label{fig:u-net}
	\vspace{-0.4cm}
\end{figure}

We compare our method with a deep ISP that adopts a widely used U-Net architecture~\cite{chen2018learning}. 
U-Net can be treated as an upper bound as it is a pure deep network solution with around 7 million tunable parameters, whilst ours involves traditional algorithms and the number of tunable parameters is just 226, with 54 architecture parameters and 172 algorithm parameters. 
In Table~\ref{tab:train_data}, we present the quantitative results with different number of training patches.\footnote{Each training patch has a resolution of 192$\times$192. Data augmentation is used including cropping and flipping.} Our method is more robust than U-Net when the data size is small. The searched ISP architecture is the same as that in Fig.~\ref{fig:isp_sid_s7}, except for the case of 100 training patches, where ``WB-Quadratic'' is replaced with ``WB-Whitepatch''. When the data size becomes large, U-Net unsurprisingly outperforms the current ReconfigISP (with a pipeline length of five). But this renders an unfair comparison to our method. 
Qualitative results are shown in Fig.~\ref{fig:u-net}. With 100 training patches, U-Net outputs a noisy image with distorted color and exposure. ReconfigISP generates a better result, indicating that our method is more robust to small-scale data. 
Apart from parameter and data efficiency, ReconfigISP possesses unique advantages of modularity and interpretability, that are not available in U-Net or other deep approaches.
Moreover, U-Net cannot flexibly adapt to specific efficiency constraints as we did in Sec.~\ref{subsec:efficiency}. A redesign of the architecture is inevitable.

\section{Discussion and Conclusion}

Tuning an ISP pipeline is notoriously labourious.
In this paper, we devise a reconfigurable ISP (ReconfigISP) that can efficiently and flexibly adjust the ISP architecture given a specific task, only by tuning hundreds of architecture weights and algorithm parameters automatically driven by designated loss functions. 
Experimental results show that ReconfigISP outperforms traditional ISP pipelines in both performance and flexibility. Notably, ReconfigISP maintains the modularity and interpretability of traditional ISP pipelines.
Thanks to this unique feature, the ISP architecture search offers additional insights for ISP design and tuning. 
For instance, deep denoising methods are more robust in RAW domain compared to sRGB domain. 
In addition, the complexity of denoising algorithms determines the ISP efficiency to a large extent. 
Lastly, white balancing contributes little to high-level tasks such as object detection.
%


\vspace{0.1cm}
\noindent
\textbf{Acknowledgement}. This study is supported under the RIE2020 Industry Alignment Fund – Industry Collaboration Projects (IAF-ICP) Funding Initiative, as well as cash and in-kind contribution from the industry partner(s).

{\small
	\vskip -1cm
    \bibliographystyle{ieee_fullname}
    \bibliography{short,ispas}
}

\end{document}